\documentclass[prd,preprint,a4paper,superscriptaddress,nofootinbib,11pt]{revtex4}
\usepackage{amsmath,amsfonts,amssymb}
\usepackage{txfonts}
\usepackage[T1]{fontenc}

\begin{document}


\begin{center}
{\bf  \Large Effects of Noncommutativity on the Black Hole Entropy}
 
 \bigskip
\bigskip
Kumar S. Gupta  {\footnote{e-mail: kumars.gupta@saha.ac.in  }}, \\  
Theory Division, Saha Institute of Nuclear Physics,
1/AF Bidhannagar, Kolkata 700 064,
India \\[3mm] 

E. Harikumar  {\footnote{e-mail: harisp@uohyd.ernet.in }}, \\  
School of Physics, University of Hyderabad,
Central University P O, Hyderabad-500046,
India \\[3mm]

Tajron Juri\'c {\footnote{e-mail: tjuric@irb.hr}}, Stjepan Meljanac {\footnote{e-mail: meljanac@irb.hr}}, Andjelo Samsarov {\footnote{e-mail: asamsarov@irb.hr}},\\  
Rudjer Bo\v{s}kovi\'c Institute, Bijeni\v cka  c.54, HR-10002 Zagreb,
Croatia \\[3mm]

\end{center}
\setcounter{page}{1}
\bigskip


\begin{center}
{\bf   Abstract}
\end{center}

In this paper the BTZ black hole geometry is probed with a noncommutative scalar field which obeys the 
$\kappa$-Minkowski algebra. The entropy of the BTZ black hole is calculated using the brick wall method. The contribution of the noncommutativity to the black hole entropy is explicitly evaluated up to the first order in the deformation parameter.  We also 
argue that such a correction to the black hole entropy can be interpreted as arising from the renormalization of the 
Newton's constant due to the effects of the noncommutativity.\\



\newpage

\section{Introduction}

Attempts to construct a quantum theory of gravity have a long history. At a macroscopic level, Einstein's theory of 
general relativity describes gravity in terms of spacetime geometry.  It is therefore natural to expect that a 
quantum theory of gravity would lead to the modification of the classical spacetime structure and that the description of spacetime as a smooth manifold may no longer be valid at the Planck scale.
Indeed, one of the approaches to quantum gravity takes this route and the spacetime is assumed to be 
noncommutative(NC) at the microscopic level \cite{cones}. Such an assumption is not ad hoc, since general relativity and 
Heisenberg's uncertainty principle together imply that the spacetime has a noncommutative structure \cite{dop1,dop2}. With this motivation, different type of noncommutative space-times and their implications to physical models have been analyzed in recent times \cite{RevNC}. 

Study of black hole physics plays an important role in exploring various quantum aspects of gravity. Even though 
black holes arose from the solutions of classical general relativity, many insights on the problem of semiclassical 
or quantum description of gravity were obtained from the study of field theories in black hole backgrounds. In 
particular, the aspect of black hole entropy and related thermodynamic properties 
\cite{bek1,bek2,hawking,unruh,gthooft} have been extensively studied in various frameworks including string 
theory \cite{ooguri,asen}, loop gravity\cite{kaul}, conformal field theory \cite{carlip,sen1,sen2} and some  related 
approaches \cite{LopezDominguez:2006wd, LopezDominguez:2009ue}. In the same spirit, there have been various attempts to construct noncommutative theories of gravity, noncommutative black hole solutions and noncommutative quantum cosmology \cite{wess1,wess2,ohl1,ohl2, mofat, ncg1, ncg2, ncg3, vor, ber1, ber2, ber3}. In particular, 
it has been shown that the noncommutative version of the BTZ black hole is described by a $\kappa$-deformed algebra 
\cite{brian1,brian2}. Similar $\kappa$-deformed algebras have been found in the noncommutative description of Kerr 
black holes \cite{schupp} and certain noncommutative versions of cosmology \cite{ohl2}. It thus appears that there 
is a certain element of universality in the appearance of the $\kappa$-deformed algebras, as they occur in the 
noncommutative(NC) descriptions of various types of classical geometries. It is therefore interesting to study 
the properties of black holes in the framework of  $\kappa$-deformed noncommutative systems. 

The $\kappa$-deformed Minkowski spacetime is defined by the algebra
\begin{equation}
[ {\hat x}_\mu, {\hat x}_\nu]=i(a_\mu {\hat x}_\nu-a_\nu{\hat x}_\mu).
\end{equation}
Here, $a_\mu$ have dimensions of length and we choose $a_0=\frac{1}{\kappa}\equiv a$ and $a_i=0$ in the later part of this paper. Note that the rhs of $\kappa$-algebra is not a constant and is more general than that of the Moyal algebra. The 
symmetry algebra of this spacetime is known as $\kappa$-Poincar\'e-Hopf algebra \cite{luk}. Various aspects of this Hopf algebra have been studied in \cite{majid, kp1,sm1,sm2,sm3}. Each realization of the $\kappa$-algebra leads to a star product which can be used to construct the twisted coproduct, which ensures the invariance under twisted diffeomorphisms.   Klein-Gordon theory in the $\kappa$-deformed spacetime was constructed \cite{epjc, andelo} and it was shown that the underlying Hopf algebra  structure leads to twisted statistics \cite{ktwist,ktwist1}. The changes due to this twisted statistics of 
Klein-Gordon field near the vicinity of black hole was analyzed in \cite{ksg}. The modification to Unruh effect due 
to $\kappa$-deformation of the spacetime was studied in \cite{rv1, rv}. Implication of the $\kappa$ deformation on 
electrodynamics  were investigated in \cite{kelec, kelec1}, and deformed geodesic equation was obtained in \cite{our}. In 
these papers, the approach adopted was to map the coordinates of the $\kappa$-deformed spacetime to that of 
commutative spacetime and using this map, functions of noncommutative coordinates were expressed 
in terms of commutative coordinates and their derivatives, as a perturbative 
expansion in powers of the deformation parameter. In these works, different realizations of mappings between 
noncommutative and commutative coordinates were used. This was done by embedding the $\kappa$-Minkowski spacetime 
algebra into Heisenberg algebra \cite{platwist, twists}. In this paper, we study 
the change in  the entropy of the BTZ black hole due to the $\kappa$-deformation of the spacetime. This is done by analyzing 
the $\kappa$-deformed Klein-Gordon field theory in the BTZ black hole back ground.

In \cite{bek1,bek2}, a parallel between properties of black holes and thermodynamic variables in $3+1$ 
dimensions were obtained. By analyzing quantum field theory in the black hole background, Hawking showed that  
black holes emit thermal radiation and further the area law of black holes was derived \cite{hawking}. Using these results and applying well known notions of quantum mechanics, 't Hooft had shown that  
there is a  divergence in the allowed energy levels of a quantum mechanical particle near the black hole horizon. But 
the gravitational effects close to the horizon of the black hole would modify the particle wave functions near the 
vicinity of the horizon and thus it is possible that these divergences would be removed. This had been modeled in 
\cite{gthooft} by introducing a cut-off in the number of allowed energy levels of the quantum 
field near the horizon of the black hole. This approach of introducing a brick wall cut-off method has also been employed in 
dimensions other than $3+1$ \cite{mann,m1}, in order to calculate the thermodynamic quantities of interest. 
 It was shown that the cut-off depends only on the black hole horizon in $3+1$ dimensions while it depends on the 
mass of the 
black hole as well as the mass of the quantum field in other dimensions \cite{mann}. It was also shown that the divergence in 
the entropy per unit area of a Klein-Gordon field propagating in the black hole background can be absorbed by 
renormalizing the gravitational constant \cite{suskind,meyers}.

In \cite{kim}, 't Hooft's approach of brick wall method was employed to calculate the entropy of BTZ black hole. The 
BTZ black hole is a solution to gravity in $2+1$ dimensions \cite{btz}. Gravity in $2+1$ dimensions has no 
propagating degrees of freedom like the $1+1$ dimensional models \cite{kim13} and thus provides a good testing ground 
for analyzing various aspects of quantum gravity and black hole physics. In \cite{kim}, entropy of a scalar theory in 
the background of BTZ metric was calculated using the brick wall cut-off, with appropriate choice of cut-off 
parameter, in the background of rotating as well as stationary BTZ black hole metrics. It was shown that the entropy 
in both cases is related to the perimeter of the BTZ black hole.  Further, this cut-off is independent of the mass 
and angular momentum of the BTZ black hole.

In this paper, we probe the geometry of a BTZ black hole using a $\kappa$-deformed noncommutative scalar field as a 
simple probe. Using the realization of the map used in \cite{kelec1, our}, we obtain the thermodynamic properties of 
the BTZ black hole in $\kappa$-deformed spacetime, by 
analyzing the $\kappa$-deformed Klein-Gordon field in the background of the BTZ black hole. 
Following \cite{kim, gthooft}, 
we calculate the entropy of the BTZ black hole in the $\kappa$-deformed spacetime by analyzing the $\kappa$-deformed 
Klein-Gordon field theory near the vicinity of the black hole horizon. Using the ideas developed in
\cite{sm1,sm2, epjc, kelec1, our}, we first obtain the Klein-Gordon theory in the $2+1$ dimensional $\kappa$-deformed 
spacetime. 
Starting with the action of this model in the $\kappa$-deformed BTZ background, we calculate the free energy and 
entropy, semi-classically using the brick wall cut-off. We restrict our attention here only to the case of 
non-rotating BTZ black hole. We start with the Klein-Gordon theory in the $\kappa$-deformed BTZ background, 
keeping terms up to leading order in the deformation parameter.  The corresponding action is shown to have higher 
derivative terms. Using the methods of higher derivative theories, we obtain the equations of motion for the scalar 
theory, 
which is valid up to the first order in the deformation parameter.  Using the WKB method, we then calculate the
energy eigenvalues of the scalar field quanta.  Using this, we calculate the free energy and entropy of the system 
where we use the brick wall cut-off.  We obtain the modification to the entropy due to the $\kappa$-deformation, 
valid up to first order in the deformation parameter $a$. This modification can be interpreted as renormalization 
of Newton's constant $G$. Noncommutativity in its most general form  is introduced and presented in Appendix A. There we use a commutator between NC curved coordinates $\hat{X}_{\mu}$ and momenta $\hat{P}_{\mu}$ to define the NC metric. As a next step we define the general NC action. We also find  the corresponding generalized equations of motion. They are valid for a generic NC scalar field $\hat{\phi}$ on a generic NC curved spacetime $\hat{g}_{\mu\nu}$. The procedure carried out in section II is actually a special case ($\mathcal{A}\neq 0$ and $\mathcal{B}=0$) of this general procedure.  We emphasize that in section II we used for simplicity  BTZ as a toy model and that the procedure presented in Appendix A and its special case of section II  are valid for a general metric $g_{\mu\nu}$.

This paper is organized as follows. In the next section, we discuss the construction of the action describing 
$\kappa$-deformed scalar theory with non-trivial metric background. Here, we use the $\star$-product for defining the 
action for $\kappa$-deformed scalar theory. We also derive the corresponding equations 
of motion  in this section.  A more general construction is discussed in appendix A. In section III, we investigate 
$\kappa$-deformed scalar theory in the background of BTZ black hole. We calculate the entropy of the BTZ black hole using 
brick-wall cut-off method. Further, we show that the modification of the entropy of the black hole due to non-commutativity 
of the spacetime can be absorbed by renormalizing the Newton's constant.  We present our concluding remarks 
and discussions in section IV. In appendix A, we present the general construction of the action describing 
scalar theory in the $\kappa$-deformed spacetime with non-trivial curvature.


\section{NC scalar field in  curved spacetime}
In this section we investigate $\kappa$-deformed scalar field in classical curved background. In the undeformed 
(commutative) spacetime, 
the action for scalar (Klein-Gordon (KG)) field is given by
\begin{equation}\label{S0}
\mathcal{S}_{0}=\int \text{d}^{4}x\sqrt{-g}\left(g^{\mu\nu}\partial_{\mu}\phi\partial_{\nu}\phi-m^{2}\phi^{2}-\xi R\phi^{2}\right)
\end{equation}
and the corresponding equation of motion is
\begin{equation}\label{eom0}
\frac{1}{\sqrt{-g}}\partial_{\mu}\left(\sqrt{-g}\ \ g^{\mu\nu}\partial_{\nu}\phi\right)+m^{2}\phi+\xi R\phi=0,
\end{equation}
where $m$ is the mass of the field, $\xi$ is a parameter and $R$ is the Ricci scalar. 
Here we study the generalization of Eqn. (\ref{S0}) and Eqn. (\ref{eom0}) to NC spacetime, more specifically to $
\kappa$-Minkowski spacetime.

The most natural (and easiest) way to do this is to promote the pointwise multiplication to a star multiplication 
in Eqn. (\ref{S0}), i.e., $f(x)g(x)\rightarrow f(x)\star g(x)$. 
There exists an isomorphism between NC algebra $\cal \hat{A}$, generated by noncommutative coordinates 
$\hat{x}_{\mu}$ and star algebra $\cal A^{\star}$, generated by commutative coordinates  
$x_{\mu}$, but with $\star$ as the algebra multiplication. Star product  between any two elements $f(x)$ and $g(x)$ 
in $\cal A^{\star}$ is defined as
\begin{equation}\label{star}
f(x)\star g(x)=\hat{f}(\hat{x})\hat{g}(\hat{x})\triangleright 1,
\end{equation}
where $\hat{f}(\hat{x})$ and $\hat{g}(\hat{x})$ are elements of $\cal \hat{A}$, and the action 
$\triangleright: \mathcal{H}\mapsto\mathcal{A}$ is defined by
\begin{equation}\label{djelovanje}
x_{\mu} \triangleright f(x)=x_{\mu}f(x),\quad p_{\mu}\triangleright f(x)=i\frac{\partial f}{\partial x^{\mu}}.
\end{equation} 
Here, $x_{\mu}$ and $p_{\mu}$ are generators of the Heisenberg algebra $\mathcal{H}$  satisfying the relations,
\begin{equation}\label{H}
[x_{\mu},x_{\nu}]=[p_{\mu},p_{\nu}]=0, \quad [p_{\mu},x_{\nu}]=i\eta_{\mu\nu},
\end{equation}
where $\eta_{\mu\nu}=\text{diag}(+,-,-,-)$ (see \cite{twists} for details on the connection between realizations, 
Heisenberg algebra and star product).
For simplicity we  consider the case where $m=\xi=0$. Thus, we postulate the following NC action 
$\hat{\mathcal{S}}$ for the scalar theory in the $\kappa$-Minkowski spacetime 
\begin{equation}\begin{split}\label{S1}
\hat{\mathcal{S}}&=\int \text{d}^{4}x\sqrt{-g}\ \ g^{\mu\nu}\left(\partial_{\mu}\phi\star\partial_{\nu}\phi\right)\\
&=\int \text{d}^{4}x\sqrt{-g}\ \ g^{\mu\nu}\left(\partial_{\mu}\hat{\phi}\partial_{\nu}\hat{\phi}\triangleright 1\right).\\
\end{split}\end{equation}
Note that the gravity is treated here as classical, i. e.  the metric $g_{\mu\nu}$ in the above is 
undeformed
\footnote{The procedure undertaken in this section is actually a special case ($\mathcal{A}\neq 0$ and $\mathcal{B}=0$), as described at the end of Appendix A. This special case consists of taking the limit of small curvatures ((A6)$\rightarrow\hat{g}_{\mu\nu}=g_{\mu\nu}+O(a\cdot \partial g)$) and of keeping only  terms linear  in the deformation parameter $a_{\mu}$.This means  that  in this section we investigate NC field $\hat{\phi}$ on the generic classical background $g_{\mu\nu}$. Hence, in the current section we take this as our starting point when deriving (17) and (19), since our aim is to investigate the leading corrections to the entropy of BTZ, originating in noncommutativity. We emphasize that for simplicity  we use here BTZ as a toy model and that the procedure presented here (as well as the procedure presented in Appendix A, which is more general one) is valid for a general metric $g_{\mu\nu}$.}.
In the case of $\kappa$-Minkowski spacetime we can use  ``theory of realizations'' (see \cite{twists} and references therein) and expand these star 
products as the power series in the deformation parameter $a_{\mu}$.

To explicitly construct the action given in Eqn. \eqref{S1} , we first need to obtain the $\star$-product defined in Eqn. 
\eqref{star}. With this in mind, we start with the $\kappa$-Minkowski space defined by
\begin{equation}\label{Minkowski}
[\hat{x}_{\mu},\hat{x}_{\nu}]=i(a_{\mu}\hat{x}_{\nu}-a_{\nu}\hat{x}_{\mu}).
\end{equation}
Operators $\hat{x}_{\mu}$ can be realized in terms of the operators $x_{\mu}$ and $p_{\mu} (=i\partial_{\mu})$ 
\cite{kelec1, our, twists} defined in the commutative spacetime as
\begin{equation}
\hat{x}_{\mu}=x_{\alpha}\varphi^{\alpha} \! _{\mu} (p).
\end{equation}
Demanding consistency of this realization with Eqn.(\ref{Minkowski}) shows that  $\varphi^{\alpha} \! _{\mu} (p)$ 
must satisfy the following conditions,
\begin{equation}\label{eqphi}
\frac{\partial \varphi^{\alpha} \! _{\mu}}{\partial p^{\beta}}\varphi^{\beta} \! _{\nu}-\frac{\partial 
\varphi^{\alpha} \! _{\nu}}{\partial p^{\beta}}\varphi^{\beta} \! _{\mu} =
a_{\mu}\varphi^{\alpha} \! _{\nu}-a_{\nu}\varphi^{\alpha} \! _{\mu}.
\end{equation}
We solve Eqn.(\ref{eqphi}) up to the first order in deformation parameter $a$ and get
\begin{equation}\label{phi}
\varphi^{\alpha} \! _{\mu}=\delta^{\alpha} _{\mu}[1+\alpha(a\cdot p)]+\beta a^{\alpha} p_{\mu}+\gamma 
p^{\alpha} a_{\mu}, \qquad  \alpha,\beta,\gamma \in \mathbb{R},
\end{equation}
where the parameters appearing in the realization, namely,  $\alpha$, $\beta$ and  $\gamma$ have to satisfy the 
constraint
\begin{equation}
\gamma-\alpha=1, \qquad \beta \in \mathbb{R}.
\end{equation}
This solution exhausts all possible covariant realizations up to the first order in deformation parameter $a$.
The realization for an arbitrary element of $\hat{\mathcal{A}}$, i.e. $\hat{f}$ is given by (see also \cite{our})
\begin{equation}
\hat{f}=f(x)+\alpha(x\cdot \frac{\partial f}{\partial x})(a\cdot p)+\beta(a\cdot x)(\frac{\partial f}{\partial x}\cdot p)+\gamma(a\cdot \frac{\partial f}{\partial x})(x\cdot p)
\end{equation}
which, up to the first order in $a$, for the star product \eqref{star} yields
\begin{equation}\label{star1}
f(x)\star g(x)=f(x)g(x)+i\alpha(x\cdot \frac{\partial f}{\partial x})(a\cdot \frac{\partial g}{\partial x})+i\beta(a\cdot x)(\frac{\partial f}{\partial x}\cdot \frac{\partial g}{\partial x})+i\gamma(a\cdot \frac{\partial f}{\partial x})(x\cdot \frac{\partial g}{\partial x})
\end{equation}
Setting $f=g=\partial\phi$ in Eqn. \eqref{S1} and Eqn. \eqref{star1}, we expand the action up to the first order 
in the deformation parameter $a_{\mu}$ as
\begin{equation}\begin{split}
\hat{\mathcal{S}}&=\mathcal{S}_{0}+\int\text{d}^{4}x\sqrt{-g}\ \ g^{\mu\nu}\left[i\alpha x^{\sigma}\frac{\partial^{2}\phi}{\partial x^{\sigma}\partial x^{\mu}}a^{\beta}+i\beta(a\cdot x)\frac{\partial^{2}\phi}{\partial x_{\beta}\partial x^{\mu}}+i\gamma\frac{\partial^{2}\phi}{\partial x_{\alpha}\partial x^{\mu}}a_{\alpha}x^{\beta}\right](\partial_{\beta}\partial_{\nu}\phi)\\
\end{split}\end{equation}
By defining,
\begin{equation}\label{A}
\mathcal{A}^{\alpha\beta\gamma\delta}=i\sqrt{-g}\ g^{\beta\delta}\left(\alpha x^{\alpha}a^{\gamma}+\beta(a\cdot x)\eta^{\alpha\gamma}+\gamma a^{\alpha}x^{\gamma}\right)
\end{equation}
we re-write the above action in a more compact form as
\begin{equation}\label{akcija}
\hat{\mathcal{S}}=\mathcal{S}_{0}+\int\text{d}^{4}x\left(\mathcal{A}^{\alpha\beta\gamma\delta}\frac{\partial^{2}\phi}{\partial x^{\alpha} \partial x^{\beta}}\frac{\partial^{2}\phi}{\partial x^{\gamma} \partial x^{\delta}}\right).
\end{equation}

Starting from the noncommutative scalar theory described by the above action we derive equations of motion for the 
field $\phi$. Notice that the action in Eqn.(\ref{akcija}) has terms involving higher derivatives of the scalar 
field, i.e., our Lagrangian is $\mathcal{L}=\mathcal{L}(\phi,\partial \phi, \partial^{2}\phi, x)$  and hence 
Euler-Lagrange equations will be more general as in the case of higher derivative theories. Thus the 
Euler-Lagrange equation relevant here is,
\begin{equation}
\partial_{\mu}\frac{\delta \mathcal{L}}{\delta(\partial_{\mu}\phi)}-\partial_{\mu}\partial_{\nu}\frac{\delta \mathcal{L}}{\delta(\partial_{\mu}\partial_{\nu}\phi)}=\frac{\delta \mathcal{L}}{\delta \phi}.
\end{equation}
Using this  we find the Euler-Lagrange equation following from Eqn.(\ref{akcija}) explicitly as
\begin{equation}\begin{split}\label{eomfull}
\partial_{\sigma}(\sqrt{-g}\ g^{\sigma\nu}\partial_{\nu}\phi)&=\partial_{\alpha}\partial_{\beta}(\mathcal{A}^{\alpha\beta\gamma\delta} \partial_{\gamma}\partial_{\delta}\phi)+\partial_{\gamma}\partial_{\delta}(\mathcal{A}^{\alpha\beta\gamma\delta}\partial_{\alpha}\partial_{\beta}\phi)\\
& -\frac{1}{2}\sum_{\alpha}\partial_{\alpha}\partial_{\alpha}(\mathcal{A}^{\alpha\alpha\gamma\delta}\partial_{\gamma}\partial_{\delta}\phi+\mathcal{A}^{\gamma\delta\alpha\alpha}\partial_{\gamma}\partial_{\delta}\phi).
\end{split}\end{equation}

\section{$\kappa$-deformed scalar theory  in the BTZ background}

So far our analysis was carried out for a general curved spacetime metric $g_{\mu\nu}(x)$. Since, NC effects are 
related to Planck scale physics, we expect that the spacetime of black holes is a natural arena for studying NC 
theories. 
Having this in mind  we use the BTZ metric \cite{btz, BTZ} explicitly in Eqn.(\ref{akcija}) and use the $\kappa$-deformed
scalar field to probe the BTZ  geometry in order to infer new features that the noncommutativity brings into the black hole physics. The 
BTZ black hole is described by the metric
\begin{equation}\label{btzmetric}
g_{\mu\nu}=\begin{pmatrix}
\frac{r^2}{l^2}-8GM&0&0\\
0&-\frac{1}{\frac{r^2}{l^2}-8GM}&0\\
0&0&-r^2\\
\end{pmatrix},
\end{equation}
where we have taken  the angular momentum to be zero, i.e.  
$J=0$. 
As said earlier,  we consider NC field 
$\hat{\phi}$ on  the undeformed background $g_{\mu\nu}$ (we use the above metric in  Eqns. 
(\ref{A}) and (\ref{akcija}) and  also choose  $a_{\mu}=(a, \vec{0})$ in what follows).
Even with these simplifying assumptions, the equations of motion are still non-trivial and thus we 
are forced to use further, physically motivated approximations. The first approximation that we take is the 
long wavelength limit, where we keep terms in the equations of motion that are of the lowest order in derivatives 
($\partial\phi>>\partial^2\phi,\partial^3\phi,\partial^4\phi$). In this approximation the terms dependent on $\alpha$ 
and $\gamma$  do not contribute since they are proportional to $\partial^{(2,3,4)}\phi$, and only terms proportional 
to  $\partial\phi$ survive. Thus, only terms depending on $\beta$
  give rise to noncommutative contributions.
Note here   that only realizations characterized with parameter $\beta$ contribute in the lowest order in the long 
wavelength approximation. The choice of realization corresponds to the choice of the vacuum of the theory and this 
should be fixed by experiment in principle. For example $\beta=1$ corresponds to the natural realization (classical 
basis \cite{twists}).  
The equation of motion is still complicated, so we solve it using WKB approximation and obtain the spectrum. We use the 
ansatz  $\phi(r,\theta,t)=R(r)e^{-i\omega t}e^{im\theta}$,  as long as $M>>1$ and keep terms up to first order in
the deformation parameter $a$. Taking all above into account, we get the radial equation as
\begin{equation}\label{eomradial}
r\left(8GM-\frac{r^2}{l^2}\right)\frac{\partial^2 R}{\partial r^2}+\left(8GM-\frac{3r^2}{l^2}\right)
\frac{\partial R}{\partial r}+\left(\frac{m^2}{r}-\omega^2\frac{r}{\frac{r^2}{l^2}-8GM}-a\beta\omega\frac{8r}{l^2}\frac{\frac{3r^2}{2l^2}-8GM}{\frac{r^2}{l^2}-8GM}\right)R=0
\end{equation} 
which will be the cornerstone of the whole subsequent analyzes, presented in this paper.

\subsection{Brick wall model and the entropy}
The calculation of entropy of the black holes using ``brick-wall model'' was introduced in \cite{gthooft} for the 
general case, and in \cite{kim},  
for the BTZ case. We are following the same line of arguments as in \cite{gthooft, kim} and consequently find from Eqn. \eqref{eomradial} that  the $r$-dependent radial wave 
number has the following form
\begin{equation}\label{k}
k^2 (r,m,\omega)=-\frac{m^2}{r^2\left(\frac{r^2}{l^2}-8GM\right)}+\omega^2\frac{1}{\left(\frac{r^2}{l^2}-8GM\right)^2}+a\beta\omega\frac{8}{l^2}\frac{\frac{3r^2}{2l^2}-8GM}{\left(\frac{r^2}{l^2}-8GM\right)^2},
\end{equation}
where we used the ansatz $R(r)=\text{e}^{i\int k(r)\text{d}r}$ and WKB approximation. According to the semi-classical quantization rule, the radial wave number is quantized as
\begin{equation}
\pi n=\int^{L}_{r_{+}+h} k(r,m,\omega)\text{d}r
\end{equation}
where the quantum number $n>0$, $m$ should be fixed such that  $k(r,m,\omega)$ is real and $h$ and $L$ are ultraviolet
and infrared regulators\footnote{In the subsequent calculation for free energy and entropy we take the limit
$L\rightarrow\infty$ and set $h\approx 0$ and we keep only the most divergent terms in $h$.}, respectively. The total 
number $\nu$ of solutions with energy not exceeding $\omega$ is given by
\begin{equation}
\nu=\sum^{m_{0}}_{-m_{0}}n=\int^{m_{0}}_{-m_{0}}\text{d}m ~n=\frac{1}{\pi}\int^{m_{0}}_{-m_{0}}\text{d}m\int^{L}_{r_{+}+h} k(r,m,\omega)\text{d}r.
\end{equation}
The free energy at inverse temperature $\beta_{T}$ of the black hole is 
\begin{equation}\begin{split}
\text{e}^{-\beta_{T}F}&=\sum_{\nu}\text{e}^{-\beta_{T}E}=\prod_{\nu}\frac{1}{1-\text{e}^{-\beta_{T}E}} \ \ \ \ \ /\ \text{ln}\\
\beta_{T}F&=\sum_{\nu}\text{ln}\left(1-\text{e}^{-\beta_{T}E}\right)=\int\text{d}\nu\text{ln}\left(1-\text{e}^{-\beta_{T}E}\right)\ \ \ \ / \  \text{part. integ.}\\
&=-\int^{\infty}_{0}\text{d}E\frac{\beta_{T}\nu(E)}{\text{e}^{\beta_{T}E}-1}
\end{split}\end{equation}
For this, we find the free energy $F$ as
\begin{equation}
F=-\frac{1}{\pi}\int^{\infty}_{0}\frac{\text{d}\omega}{\text{e}^{\beta_{T}\omega}-1}\int^{L}_{r_{+}+h}\text{d}r\int^{m_{0}}_{-m_{0}}\text{d}m\ \ k(r,m,\omega).
\end{equation}
After carrying out the integrations and keeping the most divergent terms in $h$, we have
\begin{equation}
F=-\frac{l^{\frac{5}{2}}}{(8GM)^{\frac{1}{4}}}\frac{\zeta(3)}{\beta^3_{T}}\frac{1}{\sqrt{2h}}-2a\beta\frac{(8GM)^{\frac{3}{4}}\sqrt{l}}{\sqrt{2h}}\frac{\zeta(2)}{\beta^2_{T}},
\end{equation}
which is the exact result in the sense of the  WKB method and $\zeta$ is the  Euler-Riemann zeta function.

Now we can evaluate the entropy for the NC massless scalar field  using the relation 
$S=\beta^2_{T}\frac{\partial F}{\partial \beta_{T}}$. Thus we get
\begin{equation}\begin{split}
S&=3\frac{l^{\frac{5}{2}}}{(8GM)^{\frac{1}{4}}}\frac{\zeta(3)}{\beta^2_{T}}\frac{1}{\sqrt{2h}}+4a\beta\frac{(8GM)^{\frac{3}{4}}\sqrt{l}}{\sqrt{2h}}\frac{\zeta(2)}{\beta_{T}}\\
&=S_{0}\left(1+\frac{4}{3}a\beta\frac{8GM}{l^2}\frac{\zeta(2)}{\zeta(3)}\beta_{T}\right),
\end{split}\end{equation}
where $S_{0}$ is the undeformed entropy for BTZ at the Hawking temperature $\beta_{T}=\frac{2\pi l^2}{r_{+}}$. This 
entropy is equivalent to the Beckenstein-Hawking entropy $S_{0}=\frac{A}{4G}=\frac{2\pi r_{+}}{4G}$. We use this equivalence to fix the cutoff $h$ as  
\begin{equation}
h=\frac{9G^2\zeta^2 (3)\sqrt{8GM}}{8l\pi^6}.
\end{equation} 
In other words we choose $h$ so that the entropy satisfies the area(perimeter) law, as was done in \cite{gthooft}.

The above result can be used to obtain a renormalization of the Newton's constant, which is plausible due to quantum effects at the Planck scale \cite{suskind, meyers}, where noncommutative effects are expected to be important. Let the Newton's constant at the Planck scale be denoted by $G^*$. Assuming that the black hole area law is satisfied with the renormalized Newton's constant $G^*$, we get
\begin{equation}\label{area}
S=\frac{A}{4G^*}
\end{equation}
This leads to the renormalized Newton's constant  $G^*$  defined by
\begin{equation}
\frac{1}{G^*}=\frac{1}{G}\left(1+\frac{8}{3}\frac{a\beta\pi}{l}\frac{\zeta(2)}{\zeta(3)}\sqrt{8GM}\right).
\end{equation}
Note that $G^*$  depends on the mass $M$ apart from $G$ and $a$ and reduces to $G$ when the noncommutative 
parameter $a=0$.

That the final result nevertheless reproduces the area law \eqref{area} does not
look very surprising, since indeed, the area law appears to be
robust and is reproduced in various modifications (invariant or not under
the local Lorentz transformations) of the field propagation (see for instance
\cite{nesterov}).

\section{Final remarks}

In this paper we have used a $\kappa$-Minkowski type noncommutative scalar field as a probe to study the BTZ black 
hole geometry. Using the brick wall method of \cite{gthooft}, we have obtained the noncommutative corrections to the 
entropy of BTZ black holes. Our results reduce to the ones known in the literature for the pure BTZ black hole 
\cite{kim} when the noncommutative parameter $a$ is set to zero, as required. 
 
Noncommutative effects are expected to be important at the Planck scale. It has been suggested that the Newton's 
constant may be renormalized due to quantum effects at the Planck scale \cite{suskind,meyers}. Within our framework, 
we have calculated the noncommutative correction to the Newton's constant up to the first order in the noncommutative 
deformation parameter.

 In this paper we have found  the correction to the area law dependent on the NC parameter.  This correction scales with the Beckenstein-Hawking entropy $S_{BH}$ (we denote it with $S_0.$). This is unlike the situation which occurs  in \cite{LopezDominguez:2006wd}, where to first order in the deformation, the noncommutative correction to the entropy has two contributions, with only one among them scaling with the Beckenstein-Hawking entropy. Moreover, this contribution that is proportional to $S_{BH}$  has the negative  sign, while the sign in our result depends on the $a\beta$. This means that the effect of noncommutativity in Ref.\cite{LopezDominguez:2006wd} is to reduce the number of microstates and lower the entropy and  in our case noncommutativity depending on the sign of $a\beta$ will increase/decrees the number of microstates and entropy. 

As it stands, our result apparently depends on the parameter $\beta,$ which in turn characterizes the realization of a given noncommutative algebra\footnote{It is interesting to note that in the so called \textsl{left covariant} realization (see \cite{twists} and references therein), which corresponds to Poincar\'e-Weyl symmetry, we have that $\beta=0$ which means that there is no NC contribution to the entropy in the leading order of the long-wavelength limit. In that case we should  include even higher derivative terms into consideration, which would make the subsequent analysis much more involved.}. Hence a natural question is what is the significance of the parameter $\beta$. For the $\kappa$-Minkowski space it has been  clarified that to each realization there exists a particular operator ordering prescription, star product, coalgebraic structure and a Drinfeld twist operator \cite{epjc, 2, 3, 1}. All these connections are realized through the one-one correspondences. In other words, the choice of the parameter $\beta$ is in one to one correspondence with the choice of the star product \cite{epjc}, which affects the physical consequences. It may be noted that even for Moyal spaces, the physics depends on the choice whether the Moyal product or the Voros product is used in the analysis and they lead to physically distinct predictions \cite{moyal}. Thus the parameter $\beta$, through its relation to the star product affects the physical predictions directly and empirical observations can be used to put bounds on it \cite{1}.

The system  we considered can equally be viewed as two subsystems separated by the boundary area, with black hole horizon taking the role of the boundary. Then one can  assign the reduced density matrix to each of the two subregions. This can be done by separately tracing the starting density matrix either  with respect to the degrees of freedom of the first subregion or with respect to the degrees of freedom of the second subregion.
Furthermore, if the system is to be described by a pure state, then the reduced density matrices of both subregions lead to the same entropy, which can be identified  with  the entanglement entropy \cite{Solodukhin:2011gn}.
 Consequently the entanglement entropies of both subsystems are equal. Put in other words, the entropy of the black hole is equal to the entropy of scalar field modes propagating in and out of the black hole horizon and this is what we have calculated.

In the framework of  AdS/CFT  correspondence, 
there are arguments \cite{Barbon:2008ut,Barbon:2008sr} pointing toward  the entanglement entropy  as being a probe for testing the nonlocal characteristics of the theory considered. Related to AdS/CFT,  two  parameters, which are dual to each other, need to be specified. These are correspondingly, the distance in the bulk $l$  and  CFT energy scale parameter $u$. Their mutual relation (UV/IR  dispersion) determines the local/nonlocal nature of the theory, with Heisenberg-like relation  $l  \sim  1/u$ being the footprint of the locality, at least in the theories where the conformal symmetry is preserved.

The authors of \cite{Barbon:2008ut} have analysed
 two particular theories  (Little String Theory (LST))
 and noncommutative Yang-Mills  (NC YM)) and by using their AdS/CFT description,  have  encountered a type of phase transition  in the sense that when the characteristic size $l$ of the region considered
falls well below the critical nonlocality length $l_c$, the area law becomes
 superseded by the volume law.
In the case of LST, when calculating the entropy  of the region of size $ l$, they were able to identify two regions.
In the first region (UV region), which is for  $l$ much smaller than the critical size $l_c$, they have found the volume entropy law (cubic in $l$ or alternatively, linear in  $l$, if expressed in terms of the  entropy density).  In this region  
the AdS  parameter $l$  does not scale with $u$, meaning that $l$ is independent of $u$ and the
underlying theory is nonlocal.
In the second region ( IR region), that is for  large $l,$ i.e.  $l >>  l_c$  and small  $u$ the entropy follows the area law and the UV/IR dispersion is of the Heisenberg type $l \sim 1/u$,  indicating the locality of 
the underlying theory in this region \cite{Barbon:2008ut}.
Moreover, in the context of NC YM  the UV/IR mixing   strongly affects the critical length scale $l_c$ that is governing the  area/volume law transition \cite{Barbon:2008ut}.  
In this way entanglement entropy selects between locality and nonlocality according to whether it follows area or volume law.

In our model  the characteristic size of the region considered enters  the entropy through  the  BTZ radius  $r_+$, which in turn enters the formula for entropy through the inverse temperature $\beta_T.$
However, our result for the entropy has the same form for any  $r_+$, i.e. it does not depend on the  characteristic size of the black hole.
In this sense, while reproducing the area law, we     did not come across such type of phase transition as mentioned above.
Nevertheless, the analysis presented so far naturally raises the question as to why the critical value should not also exist for $ r_+$ of BTZ. Of course, we could speculate that some additional effects enter the story as  $r_+$ approaches smaller/greater values and this would certainly be an interesting  question to pursue further.

At the end, we briefly comment that  we are using kappa-type of deformation for which UV/IR  mixing is not very well understood, certainly not as good as for Moyal case, and hence the direct comparison between the two is not easy.
The illumination of the role UV/IR mixing  has in the context of kappa-deformation and its comparison with the Moyal case would be a possible future area of work.
Furthermore,
 in the brick wall type calculations, that we carried here, the terms dependent on the infra
red cutoff are scaled away and only the most divergent term as a function
of the brick wall cutoff is retained. That is what has been done here also
and hence we do not see any infrared dependence. We can also argue that according to the arguments of  Ref.\cite{Barbon:2008ut}, our underlying theory  is local due to  entropy which follows the area law.

We have used BTZ spacetime as a model for the black hole. Note that the BTZ metric appears in the discussion of 
near-horizon geometry of a large class of black holes \cite{ooguri}. Similarly, the $\kappa$-Minkowski type of 
noncommutativity that has been 
used here appears in the noncommutative description of several black hole geometries. Thus it is plausible that our 
results have a certain amount of universality which makes it relevant for a wider class of geometries.
\bigskip

\appendix
\section{NC scalar field in NC curved spacetime}
In the absence of true NC of quantum gravity theory, we have been forced to limit our analyzes to the 
$\kappa$-deformations of fields and treat the gravity classically. However, in \cite{our}, authors have analyzed 
the most general equations of motion for particles in $\kappa$-deformed curved space. In this setting, i.e. 
``Feynman approach'', the commutator between coordinates $\hat{X}_{\mu}$ and momenta $\hat{P}_{\nu}$ can be 
interpreted as the NC metric
\begin{equation} \label{XP}
 [\hat{X}_{\mu},\hat{P}_{\nu}]\equiv -i\hat{g}_{\mu\nu}=-ig_{\alpha\beta}(\hat{y})\left(p^{\beta} \frac{\partial 
 \varphi^{\alpha} \! _{\nu}}{\partial p^{\sigma}}\varphi^{\sigma} \! _{\mu} +\varphi^{\alpha} \! _{\nu} 
 \varphi^{\beta} \! _{\mu} \right)
\end{equation}
where
\begin{equation}\label{gy}
g_{\mu\nu}(\hat{y})=g_{\mu\nu}(x)+\gamma (x\cdot \frac{\partial g_{\mu\nu}}{\partial x})(a\cdot p)+
\alpha(a\cdot\frac{\partial g_{\mu\nu}}{\partial x})(x\cdot p)+\beta(x\cdot a)(\frac{\partial g_{\mu\nu}}{\partial x}
\cdot p)
\end{equation}
Eqn. \eqref{XP} enables as to analyze $\kappa$-deformations of metric also.
This way we can postulate the following NC covariant action
\footnote{ For example in the less complicated case of NC space called Moyal space which is defined by $[\hat{x}_{\mu},\hat{x}_{\nu}]=i\Theta_{\mu\nu}$, tensor $\Theta_{\mu\nu}$ is treated as a constant tensor. In this approach (see \cite{wess1,wess2,ohl1,ohl2,schenkel} and references therein ), the symmetries of general relativity, i.e. the diffeomorphism symmetry is formulated in the language of Hopf algebras, which provide a mathematical framework suitable for studying quantization of Lie groups  and Lie algebras. A gravity theory is then constructed in such a way that it transforms covariantly under the deformed diffeomorphisms, which automatically lead to noncommutative geometry. Here it is important to introduce the notion of star product, twisted symmetries and twist operator and their mutual relations. The construction outlined in this Appendix (as well as  its  special case treated  in section II) is covariant with respect to twisted diffeomorphisms. We further emphasize that it is possible to construct the corresponding twist operator (along the line of Ref. \cite{twists}, where it is shown that for every realization there is a corresponding twist operator leading to a unique associative star product) underlying the construction presented in this Appendix  and which ensures the general covariance of the starting action. Here we have started from the star product and its realization. However, we could equally well  start from the twist and then, using the twisted diffeomorphism, obtain  the same equations of motion. 
The main advantage in using the star product is that the action written in terms of it is automatically covariant under  the  action of deformed symmetries. Of  course in the limit $a_{\mu}\rightarrow 0$ the star product reduces to commutative multiplication and we recover the usual notion of covariance.}
$\hat{\mathcal{S}}$ for the NC scalar theory in the 
$\kappa$-deformed spacetime with non-trivial metric as
\begin{equation}\begin{split}\label{S1a}
\hat{\mathcal{S}}&=\int \text{d}^{4}x\sqrt{-g}\left(g^{\mu\nu}\star\partial_{\mu}\phi\star\partial_{\nu}\phi\right)\\
&=\int \text{d}^{4}x\sqrt{-g}\left(\hat{g}^{\mu\nu}\partial_{\mu}\hat{\phi}\partial_{\nu}\hat{\phi}\triangleright 1
\right).\\
\end{split}\end{equation}
In the case of $\kappa$-Minkowski spacetime we can use  ``theory of realizations'' (see \cite{twists} and references therein) and expand these star 
products as power series in the deformation parameter $a_{\mu}$.
Note that in the above we leave the volume $d^4x$ undeformed \cite{novi}. Using the $\hat{\phi}$ and 
$\hat{g}_{\mu\nu}$  (see \cite{our} for details) valid up to first order in the deformation parameter, 
\begin{equation}\label{phii}
\hat{\phi}=\phi(x)+\alpha(x\cdot \frac{\partial \phi}{\partial x})(a\cdot p)+\beta(a\cdot x)(\frac{\partial \phi}
{\partial x}\cdot p)+\gamma(a\cdot \frac{\partial \phi}{\partial x})(x\cdot p),
\end{equation}
\begin{equation}\begin{split}\label{ghat}
\hat{g}_{\mu\nu}=&g_{\mu\nu}(\hat{y})+2\alpha g_{\mu\nu}(a\cdot p)+\alpha g_{\nu\beta}a_{\mu}p^{\beta}+\beta(\eta_{\mu\nu}g_{\alpha\beta}a^{\alpha}p^{\beta}+g_{\nu\beta}a^{\beta}p_{\mu}+g_{\mu\alpha}a^{\alpha}p_{\nu})\\
&+\gamma(g_{\mu\beta}a_{\nu}p^{\beta}+g_{\nu\beta}p^{\beta}a_{\mu}+g_{\mu\alpha}p^{\alpha}a_{\nu}),\\
\end{split}\end{equation}
respectively, we now expand the action (valid up to the first order in the deformation parameter $a_{\mu}$). 
Rewriting the above equations in a more compact form as
\begin{equation}\begin{split}\label{razvoj}
&\hat{g}_{\mu\nu}=g_{\mu\nu}+ia_{\alpha}G^{\alpha\beta} _{\mu\nu}(x)\partial_{\beta}\equiv g_{\mu\nu}+\delta(g_{\mu\nu}), \quad \delta(g_{\mu\nu})\triangleright 1=0\\
&\partial_{\mu}\hat{\phi}=\partial_{\mu}\phi+ia_{\alpha}\varphi^{\alpha\beta} _{\mu}(x)\partial_{\beta}\equiv \partial_{\mu}\phi+\delta(\partial_{\mu}\phi), \quad \delta(\partial_{\mu}\phi)\triangleright 1=0,\\
\end{split}\end{equation}
we expand the action in Eqn. (\ref{S1a}) keeping terms up to the first order in deformation parameter. Thus we find
\begin{equation}\begin{split}\label{S}
\hat{\mathcal{S}}&=\int \text{d}^{4}x\sqrt{-g}\left(g_{\mu\nu}+\delta(g_{\mu\nu})\right)\left(\partial_{\mu}\phi+\delta(\partial_{\mu}\phi)\right)\left(\partial_{\nu}\phi+\delta(\partial_{\nu}\phi)\right)\triangleright 1\\
&=\int \text{d}^{4}x\sqrt{-g}\left(g_{\mu\nu}+\delta(g_{\mu\nu})\right)\left(\partial_{\mu}\phi\partial_{\nu}\phi+\delta(\partial_{\mu}\phi)\partial_{\nu}\phi\right)\triangleright 1\\
&=\int \text{d}^{4}x\sqrt{-g}\left(g_{\mu\nu}+\delta(g_{\mu\nu})\right)\left(\partial_{\mu}\phi\partial_{\nu}\phi+[\delta(\partial_{\mu}\phi),\partial_{\nu}\phi]\right)\triangleright 1\\
&=\int \text{d}^{4}x\sqrt{-g}\left(g^{\mu\nu}\partial_{\mu}\phi\partial_{\nu}\phi+g^{\mu\nu}[\delta(\partial_{\mu}\phi),\partial_{\nu}\phi]+[\delta(g_{\mu\nu}),\partial_{\mu}\phi\partial_{\nu}\phi]\right)+O(a^2)\\
&\equiv \mathcal{S}_{0}+a\mathcal{S}'.
\end{split}\end{equation}
By using the following set of relations \footnote{note that in (\ref{gy} - \ref{A9})the $\cdot$ product is with respect to $\eta_{\mu\nu}$, so we also have $g_{\mu}^{\ \nu}\equiv g_{\mu\alpha}\eta^{\alpha\beta}$.},
\begin{subequations}
\begin{equation}
\delta(\partial_{\mu}\phi)=ia_{\alpha}\varphi^{\alpha\beta} _{\mu}(x)\partial_{\beta},
\end{equation}
with
\begin{equation}
\varphi^{\alpha\beta} _{\mu}(x)=\alpha\left(x\cdot\frac{\partial^{2}\phi}{\partial x \partial x^{\mu}}\right)\eta^{\alpha\beta}+\beta x^{\alpha}\frac{\partial^{2}\phi}{\partial x_{\beta}\partial x^{\mu}}+\gamma\frac{\partial^{2}\phi}{\partial x_{\alpha}\partial x^{\mu}}x^{\beta},
\end{equation}
and
\begin{equation}
\delta(g_{\mu\nu})=ia_{\alpha}G^{\alpha\beta} _{\mu\nu}(x)\partial_{\beta},
\end{equation}
with
\begin{equation}\begin{split}
G^{\alpha\beta} _{\mu\nu}(x)&=\gamma\left(x\cdot\frac{\partial g_{\mu\nu}}{\partial x}\right)\eta^{\alpha\beta}+\alpha\frac{\partial g_{\mu\nu}}{\partial x_{\alpha}}x^{\beta}+\beta x^{\alpha}\frac{\partial g_{\mu\nu}}{\partial x_{\beta}}+2\alpha g_{\mu\nu}\eta^{\alpha\beta}+\alpha g_{\nu}^{\ \beta}\delta_{\mu} ^{\alpha}\\
&+\beta(\eta_{\mu\nu}g_{\alpha}^{\ \beta}+g_{\nu\alpha}\delta_{\mu}^{\beta}+g_{\mu\alpha}\delta_{\nu}^{\beta})+\gamma(g_{\mu}^{\ \beta}\delta_{\nu}^{\alpha}+g_{\nu}^{\ \beta}\delta_{\mu}^{\alpha}+g_{\mu}^{\ \beta}\delta_{\nu}^{\alpha}),
\end{split}\end{equation}
\end{subequations}
we obtain the expression for the  action in Eqn.(\ref{S}) as
\begin{equation}\begin{split}\label{A9}
\hat{\mathcal{S}}&=\mathcal{S}_{0}+\int\text{d}^{4}x\sqrt{-g}\Bigg\{g^{\mu\nu}\left[i\alpha x^{\sigma}\frac{\partial^{2}\phi}{\partial x^{\sigma}\partial x^{\mu}}a^{\beta}+i\beta(a\cdot x)\frac{\partial^{2}\phi}{\partial x_{\beta}\partial x^{\mu}}+i\gamma\frac{\partial^{2}\phi}{\partial x_{\alpha}\partial x^{\mu}}a_{\alpha}x^{\beta}\right](\partial_{\mu}\partial_{\nu}\phi)\\
&+ia_{\alpha}G_{\mu\nu}^{\alpha\beta}\left[\frac{\partial^{2}\phi}{\partial x^{\beta}\partial x_{\mu}}\frac{\partial \phi}{\partial x_{\nu}}+\frac{\partial^{2}\phi}{\partial x^{\beta}\partial x_{\nu}}\frac{\partial \phi}{\partial x_{\mu}}\right] \Bigg\},
\end{split}\end{equation}
which is valid up to first order in the deformation parameter. Further, by defining
\begin{subequations}\label{AB}
\begin{equation}
\mathcal{A}^{\alpha\beta\gamma\delta}=i\sqrt{-g}\ g^{\beta\delta}\left(\alpha x^{\alpha}a^{\gamma}+\beta(a\cdot x)\eta^{\alpha\gamma}+\gamma a^{\alpha}x^{\gamma}\right)
\end{equation}
and 
\begin{equation}
\mathcal{B}^{\beta} _{\rho\sigma}=i\sqrt{-g}\ a_{\alpha}\left(G^{\alpha\beta} _{\rho\sigma}+G^{\alpha\beta} _{\sigma\rho}\right),
\end{equation}
\end{subequations}
we rewrite the above action in a more compact form as
\begin{equation}\label{akcijaa}
\hat{\mathcal{S}}=\mathcal{S}_{0}+\int\text{d}^{4}x\left(\mathcal{A}^{\alpha\beta\gamma\delta}\frac{\partial^{2}\phi}{\partial x^{\alpha} \partial x^{\beta}}\frac{\partial^{2}\phi}{\partial x^{\gamma} \partial x^{\delta}}+\mathcal{B}^{\alpha\beta\gamma}\frac{\partial^{2}\phi}{\partial x^{\alpha}\partial x^{\beta}}\frac{\partial \phi}{\partial x^{\gamma}}\right).
\end{equation}
Eqn. \eqref{akcijaa} represents the NC action for NC scalar field in NC background expanded to the first order in the deformation parameter $a$.

Starting from the noncommutative scalar field theory described by the action \eqref{akcijaa}, we derive equations of 
motion for the field $\phi$. Note that the action in Eqn.(\ref{akcijaa}) has terms involving  higher 
derivatives of the scalar field, i.e., our Lagrangian is $\mathcal{L}=\mathcal{L}(\phi,\partial \phi, 
\partial^{2}\phi, x)$. Hence, as earlier, we derive the Euler-Lagrange equation as that in 
higher derivative theories. Thus the Euler-Lagrange equation relevant here is,
\begin{equation}
\partial_{\mu}\frac{\delta \mathcal{L}}{\delta(\partial_{\mu}\phi)}-\partial_{\mu}\partial_{\nu}\frac{\delta \mathcal{L}}{\delta(\partial_{\mu}\partial_{\nu}\phi)}=\frac{\delta \mathcal{L}}{\delta \phi}.
\end{equation}
In order to calculate equations of motion we notice that,
\begin{subequations}
\begin{equation}
\frac{\delta \mathcal{L}}{\delta \phi}=0, \quad \frac{\delta(\partial_{\alpha}\phi)}{\delta(\partial_{\mu}\phi)}=\delta^{\mu} _{\alpha}, \quad \frac{\delta(\partial_{\alpha}\partial_{\beta}\phi)}{\delta(\partial_{\mu}\partial_{\nu}\phi)}=\delta^{\mu}_{\alpha}\delta^{\nu}_{\beta}+\delta^{\nu}_{\alpha}\delta^{\mu}_{\beta}-\Theta_{\alpha\beta}^{\mu\nu},
\end{equation}
\begin{equation}
\frac{\delta\mathcal{L}}{\delta(\partial_{\sigma}\phi)}=2\sqrt{-g}\ g^{\sigma}_{\ \nu}\partial^{\nu}\phi+\mathcal{B}^{\alpha\beta\sigma}(\partial_{\alpha}\partial_{\beta}\phi),
\end{equation}
and
\begin{equation}\begin{split}
\frac{\delta \mathcal{L}}{\delta (\partial_{\sigma}\partial_{\rho}\phi)}&=\mathcal{A}^{\alpha\beta\gamma\delta}\left[(\delta^{\sigma}_{\alpha}\delta^{\rho}_{\beta}+\delta^{\rho}_{\alpha}\delta^{\sigma}_{\beta}-\Theta_{\alpha\beta}^{\rho\sigma})(\partial_{\gamma}\partial_{\delta}\phi)+(\delta^{\sigma}_{\gamma}\delta^{\rho}_{\delta}+\delta^{\rho}_{\gamma}\delta^{\sigma}_{\delta}-\Theta_{\gamma\delta}^{\rho\sigma})(\partial_{\alpha}\partial_{\beta}\phi)\right]\\
&+\mathcal{B}^{\alpha\beta\gamma}(\delta^{\sigma}_{\alpha}\delta^{\rho}_{\beta}+\delta^{\rho}_{\alpha}\delta^{\sigma}_{\beta}-\Theta_{\alpha\beta}^{\rho\sigma})(\partial_{\gamma}\phi).
\end{split}\end{equation}
\end{subequations}
Using this (where we defined $\Theta_{\alpha\beta}^{\mu\nu}=1$ 
when $\alpha=\beta=\mu=\nu$ and $\Theta_{\alpha\beta}^{\mu\nu}=0$ otherwise) we find the Euler-Lagrangian equation 
explicitly as
\begin{equation}\begin{split}\label{eomfulla}
\partial_{\sigma}(\sqrt{-g}\ g^{\sigma\nu}\partial_{\nu}\phi)&=\partial_{\alpha}\partial_{\beta}(\mathcal{A}^{\alpha\beta\gamma\delta} \partial_{\gamma}\partial_{\delta}\phi)+\partial_{\gamma}\partial_{\delta}(\mathcal{A}^{\alpha\beta\gamma\delta}\partial_{\alpha}\partial_{\beta}\phi)\\
&+\partial_{\alpha}\partial_{\beta}(\mathcal{B}^{\alpha\beta\gamma}\partial_{\gamma}\phi)-\frac{1}{2}\partial_{\sigma}(\mathcal{B}^{\alpha\beta\sigma}\partial_{\alpha}\partial_{\beta}\phi)\\
& -\frac{1}{2}\sum_{\alpha}\partial_{\alpha}\partial_{\alpha}(\mathcal{A}^{\alpha\alpha\gamma\delta}\partial_{\gamma}\partial_{\delta}\phi+\mathcal{A}^{\gamma\delta\alpha\alpha}\partial_{\gamma}\partial_{\delta}\phi+\mathcal{B}^{\alpha\alpha\gamma}\partial_{\gamma}\phi).
\end{split}\end{equation}
Though, the Eqn.\eqref{eomfulla} is in principle highly nontrivial,  we can separate the analysis  in three different 
cases:
\begin{itemize}
\item NC field $\hat{\phi}$ on undeformed background $g_{\mu\nu}$, in this case we have $\mathcal{A}\neq 0$, $\mathcal{B}=0$, where $\mathcal{A}, \mathcal{B}$ are defined in Eqn. (\ref{AB}).
\item Commutative field $\phi$ on deformed background $\hat{g}_{\mu\nu}$: $\mathcal{A}=0$, $\mathcal{B}\neq 0$,
\item NC field $\hat{\phi}$ on deformed background $\hat{g}_{\mu\nu}$: $\mathcal{A}\neq 0$, $\mathcal{B}\neq 0$.
\end{itemize}
We have analyzed the  first situation in this paper. Other two cases are presently under investigation and will be 
reported elsewhere.


\end{document}